\documentclass[10pt]{article}
\usepackage[OE]{express}
\usepackage{amsmath}
\usepackage{subfigure}

\begin{document}
\title{Driving many distant atoms into high-fidelity steady state entanglement via Lyapunov control}

\author{Chuang Li,\authormark{1} Jie Song,\authormark{1,3} Yan Xia,\authormark{2} and Weiqiang Ding\authormark{1,4}}

\address{
\authormark{1}Department of Physics, Harbin Institute of Technology, Harbin, 150001, China\\
\authormark{2}Department of Physics, Fuzhou University, Fuzhou, 350002, China\\
}

\email{\authormark{3}jsong@hit.edu.cn} 
\email{\authormark{4}wqding@hit.edu.cn} %% email address is required



\begin{abstract}
Based on Lyapunov control theory in closed and open systems, we propose a scheme to generate W state of many distant atoms in the cavity-fiber-cavity system.
In the closed system, the W state is generated successfully even when the coupling strength between the cavity and fiber is extremely weak.
In the presence of atomic spontaneous emission or cavity and fiber decay, the photon-measurement and quantum feedback approaches are proposed 
to improve the fidelity, which enable efficient generation of high-fidelity W state in the case of large dissipation. Furthermore, the time-optimal Lyapunov control is investigated to shorten the evolution time and improve the fidelity in open systems.
\end{abstract}

\ocis{(270.0270) Quantum optics; (060.5565) Quantum communications.} % REPLACE WITH CORRECT OCIS CODES FOR YOUR ARTICLE, MINIMUM OF TWO; Avoid using the OCIS codes for “General” or “General science” whenever possible.
%For a complete list of OCIS codes, visit: https://www.osapublishing.org/oe/submit/ocis/

%%%%%%%%%%%%%%%%%%%%%%% References %%%%%%%%%%%%%%%%%%%%%%%%%
%\begin{thebibliography}{99}
%
%\bibitem{gallo99} K. Gallo and G. Assanto, ``All-optical diode based on second-harmonic generation in an asymmetric waveguide,'' \josab {\bfseries 16}(2), 267--269 (1999).
%
%\end{thebibliography}

\bibliographystyle{osajnl.bst}
\bibliography{reference.bib}

%%%%%%%%%%%%%%%%%%%%%%%%%%  body  %%%%%%%%%%%%%%%%%%%%%%%%%%
\section{Introduction}

Entanglement is a major resource in a vast number of applications in quantum information
\cite{PhysRevLett.67.661,PhysRevLett.70.1895,PhysRevLett.76.4656,PhysRevLett.86.5188,RevModPhys.81.865,
	1367-2630-12-2-025017,PhysRevLett.106.050501,PhysRevLett.111.010501,1367-2630-18-4-043031,PhysRevLett.118.257402}.
In particular, multipartite entanglement as a crucial role in quantum information processing, has attracted a lot of
attention
\cite{PhysRevLett.112.120505,PhysRevLett.114.113604,PhysRevLett.117.210504,PhysRevLett.117.240503,PhysRevLett.118.036102,PhysRevX.7.021042,1742-5468-2017-5-053104}.
Generation of entanglement is always a research hotspot and many different schemes have been proposed to generate entangled
states, such as trapped ions \cite{PhysRevLett.81.3631,sackett2000experimental},
quantum electrodynamics \cite{PhysRevLett.85.2392,wallraff2004strong,PhysRevA.88.062311,doi:10.1080/09500340.2015.1044761,PhysRevA.94.012302,PhysRevA.94.012309}, and photon
pairs \cite{PhysRevLett.47.460,PhysRevLett.103.020504}.
Especially, cavity QED system is a simple and efficient source for generating entanglement \cite{1555-6611-24-12-125203}.

Quantum control is an important technology in quantum information and quantum optics,  which has received
wide attention
\cite{d2007introduction,PhysRevLett.115.200502,PhysRevLett.116.143602,PhysRevA.96.023403,PhysRevA.96.012333}.
Different control strategies have been presented for realizing quantum control, such as
optimal control \cite{Dolde2014}, learning control \cite{PhysRevA.89.023402},
sliding mode control \cite{1367-2630-11-10-105033}, and Lyapunov control \cite{1367-2630-11-10-105034}.
Among these control strategies, the design process based on Lyapunov control method is simple and visualized,
and the control laws can ensure the stability of the control system, thus Lyapunov control is applied widely in quantum information processing
\cite{PhysRevA.80.052316,PhysRevA.82.034308,0953-4075-44-16-165503,6160433,PhysRevA.86.022321,ZHAO20121833,2014arXiv1401.2495W,PhysRevA.91.032301}.
The basic principle of Lyapunov control is that steering the system into the target via time-varying control fields, which are determined by
the Lyapunov function.
In the procedure of the Lyapunov control design, one first selects the Lyapunov function according to the target,
and then designs the control fields via the time derivative of the selected Lyapunov function.
Recently, many schemes of entanglement generation via Lyapunov control are proposed\cite{PhysRevA.80.042305,Shi2016}.
However, most studies of entanglement generation based on Lyapunov control are restricted to bipartite entanglement
and multipartite entanglement in a single cavity.
Multipartite entanglement generation for distant atoms can be a key breakthrough point in quantum information processing
that deserves research efforts.

In this paper, we propose a scheme for generating W state for three-level atoms trapped in distant cavities.
Based on Lyapunov control, we generate a W state of atoms in closed and open systems and discuss the effect of
system parameters on the fidelity.
In open systems, we use photon measurement and quantum feedback to improve the fidelity.
In addition, we investigate time-optimal Lyapunov control to shorten the time required to reach
the target state.
The rest of the paper is organized as follows:
In section~\ref{s1}, we introduce the model of the cavity-fiber-cavity system.
We investigate generation of W state for atoms in closed and open systems in section~\ref{s2} and \ref{s3},
respectively.
The time-optimal Lyapunov control is discussed in section~\ref{s4}.
In section~\ref{s5}, we summarize and conclude.

\section{The model of the cavity-fiber-cavity system}\label{s1}

We consider a cavity-fiber-cavity system, where two distant cavities are connected by an optical fiber.
Two cavities are resonantly coupled to a fiber mode $b$ with strength $\nu$.
The two cavities contain a three-level atom and two three-level atoms, respectively.
Each atom has an excited state $|e\rangle$ and two ground states $|g\rangle$ and $|f\rangle$.
The energy of the state $|g\rangle$ is taken to be zero as the energy reference point.
The states $|e\rangle$ and $|f\rangle$ have energies $\omega_e$ and $\omega_f$, respectively ($\hbar=1$).
The transition $|e\rangle\leftrightarrow|g\rangle$ is coupled to a cavity mode with strength $g$.
The transition $|e\rangle\leftrightarrow|f\rangle$ is driven by a classical field with strength $\Omega$.
The Hamiltonian of the cavity-fiber-cavity system contains four parts
\begin{equation}
H=H_0+H_{\mathrm{ac}}+H_{\mathrm{fc}}+H_{\mathrm{d}},
\end{equation}
with
\begin{subequations}
\begin{eqnarray}
H_0&=&\sum_{j=1,2,3}(\omega_e|e_j\rangle\langle e_j|+\omega_f|f_j\rangle\langle f_j|)
+\sum_{i=1,2}\omega_ca_i^\dag a_i+\omega_c b^{\dag}b,\\
H_{\mathrm{ac}}&=&g_1|e_1\rangle\langle g_1|a_1+\sum_{j=2,3}g_j|e_j\rangle\langle g_j|a_2+\mathrm{H.c.},\\
H_{\mathrm{fc}}&=&\nu b(a_1^{\dag}+a_2^{\dag})+\mathrm{H.c.},\\
H_{\mathrm{d}}&=&\sum_{j=1,2,3}\Omega_j\mathrm{e}^{-\mathrm{i}\omega_{l}t}(|e_j\rangle\langle f_j|+\mathrm{H.c.}),
\end{eqnarray}
\end{subequations}
where the subscripts $i=1,2$ and $j=1,2,3$ correspond to the $i$th cavity and the $j$th atom, respectively.
$a_i$ ($b$) is the annihilation operator of the cavity (fiber) mode.
$\omega_c$ is the frequency of the cavity (fiber) modes and $\omega_{l}$ is the frequency of the classical fields.
By using the transformation operator $U^{\dag}=\exp(-\mathrm{i}\Delta t)$, the Hamiltonian in interaction picture is given by
\begin{eqnarray}
H_{\mathrm{I}}&=&\sum_{j=1,2,3}\{\Delta|e_j\rangle\langle e_j|
+\Omega_j[|e_j\rangle\langle f_j|+\nu b(a_1^{\dag}+a_2^{\dag})+\mathrm{H.c.}] \}\nonumber\\
&&+(g_1|e_1\rangle\langle g_1|a_1+\sum_{j=2,3}g_j|e_j\rangle\langle g_j|a_2+\mathrm{H.c.}),
\end{eqnarray}
where $\Delta=\omega_e-\omega_f-\omega_l=\omega_e-\omega_c$.
We change the phases of classical fields in cavity $2$, which leads to $\Omega_2\rightarrow-\Omega_2$ and
$\Omega_3\rightarrow-\Omega_3$.
The correspond Hamiltonian of the system can be written as
\begin{eqnarray}
H_{\mathrm{I}}&=&\sum_{j=1,2,3}\Delta|e_j\rangle\langle e_j|
+(g_1|e_1\rangle\langle g_1|a_1+\sum_{j=2,3}g_j|e_j\rangle\langle g_j|a_2+\mathrm{H.c.})\nonumber\\
&&
+[\Omega_1|e_1\rangle\langle f_1|-\Omega_2|e_2\rangle\langle f_2|-\Omega_3|e_3\rangle\langle f_3|
+\nu b(a_1^{\dag}+a_2^{\dag})+\mathrm{H.c.}].
\end{eqnarray}
Assuming the system is initially in the state $|f,g,g,0,0,0\rangle_{1,2,3,a_1,a_2,b}$, the dark state
(the eigenstate of the Hamiltonian with zero eigenvalue) is
\begin{eqnarray}
|D\rangle&=&d_1|f,g,g,0,0,0\rangle+d_2|g,f,g,0,0,0\rangle+d_3|g,g,f,0,0,0\rangle \nonumber\\
&&+d_4|g,g,g,1,0,0\rangle+d_5|g,g,g,0,1,0\rangle,
\end{eqnarray}
where
\begin{subequations}
\begin{eqnarray}
d_1=\frac{g_1\Omega_2\Omega_3}{\sqrt{N_d}},
\quad d_2=\frac{g_2\Omega_1\Omega_3}{\sqrt{N_d}},
\quad d_3=\frac{g_3\Omega_1\Omega_2}{\sqrt{N_d}},
\quad
d_4=-d_5=-\frac{\Omega_1\Omega_2\Omega_3}{\sqrt{N_d}},\\
N_d=(g_1\Omega_2\Omega_3)^2+(g_2\Omega_1\Omega_3)^2+(g_3\Omega_1\Omega_2)^2+2(\Omega_1\Omega_2\Omega_3)^2.
\end{eqnarray}
\end{subequations}
We set $g_1=g_2=g_3=g$ and $\Omega_1=\Omega_2=\Omega_3=\Omega$.
When $g\gg\Omega$, the dark state is approximated to a W state of three atoms, i.e.,
\begin{equation}
|D\rangle\simeq|\psi_{\mathrm{T}}\rangle=\frac{1}{\sqrt{3}}(|f,g,g\rangle+|g,f,g\rangle+|g,g,f\rangle)\otimes|0,0,0\rangle.
\end{equation}
The condition $g\gg \Omega$ can be realized by choosing the parameters as $\Omega=0.1g$.
For example, if the coupling strength is chosen as $g=2\pi \times 34$ MHz \cite{PhysRevLett.93.233603}, the Rabi frequency
is about $\Omega\simeq2\pi \times 3.4$ MHz.

%%Fig. 1
\begin{figure}
	\centering
	\includegraphics[width=0.5\textwidth]{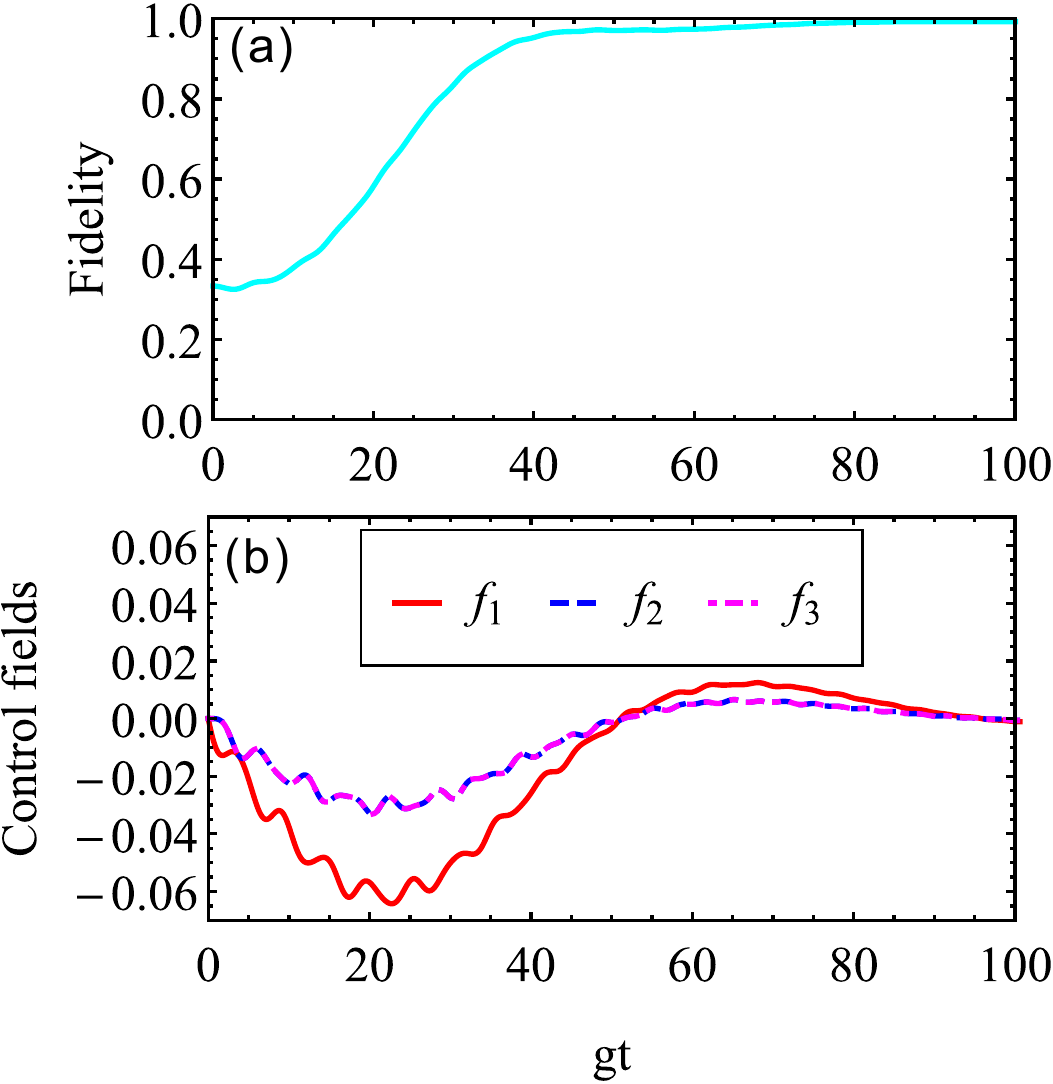}
	\caption{The time evolution of (a) the fidelity and (b) the control fields.
		Other parameters: $\Delta=0$, $\Omega=0.1g$, $\nu=g$, and $K=0.2$.\label{f3-1}}
\end{figure}

\section{Lyapunov control in closed systems}\label{s2}
In this section, we demonstrate how to steer the system into the W state $|\psi_{\mathrm{T}}\rangle$ via Lyapunov control.
In closed systems, the dynamical evolution of the system satisfies quantum Liouville equation ($\hbar=1$)
\begin{equation}
\dot{\rho}=-\mathrm{i}[H_{\mathrm{I}}+\sum_k f_k(t)H_k,\rho],
\end{equation}
where $H_k$ is the time-independent control Hamiltonian and $f_k(t)$ is time-varying control field.
The Lyapunov function is defined as
\begin{equation}
V=1-\mathrm{Tr}(\rho_{\mathrm{T}}\rho),
\end{equation}
where $\rho_{\mathrm{T}}=|\psi_{\mathrm{T}}\rangle \langle \psi_{\mathrm{T}}|$.
We calculate the time derivative of the Lyapunov function to design the control fields,
\begin{eqnarray}\label{18}
\dot{V}&=&-\mathrm{Tr}(-\mathrm{i}\rho_{\mathrm{T}}[H_{\mathrm{I}}+\sum_k f_k(t)H_k,\rho])\nonumber\\
&=&-\mathrm{Tr}(-\mathrm{i} \rho_{\mathrm{T}}[H_{\mathrm{I}},\rho])-\sum_k f_k(t)\mathrm{Tr}(-\mathrm{i}\rho_{\mathrm{T}}[H_k,\rho])\nonumber\\
&=&-\sum_kf_k(t)T_k,
\end{eqnarray}
where $T_k=\mathrm{Tr}(-\mathrm{i}\rho_{\mathrm{T}}[H_k,\rho])$.
The Lyapunov control strategy requires $\dot{V}\le0$, hence the control fields can be designed as
\begin{equation}\label{1}
f_k(t)=KT_k,
\end{equation}
where $K>0$, may be chosen properly to adjust the control amplitude.
Considering atomic spontaneous emission and convenience of experimental implementation,
we choose the control Hamiltonians as
\begin{equation}
H_k=|e_k\rangle\langle f_k|+|f_k\rangle\langle e_k|,\quad k=1,2,3.
\end{equation}

\begin{figure}
	\centering
	\includegraphics[width=0.8\textwidth]{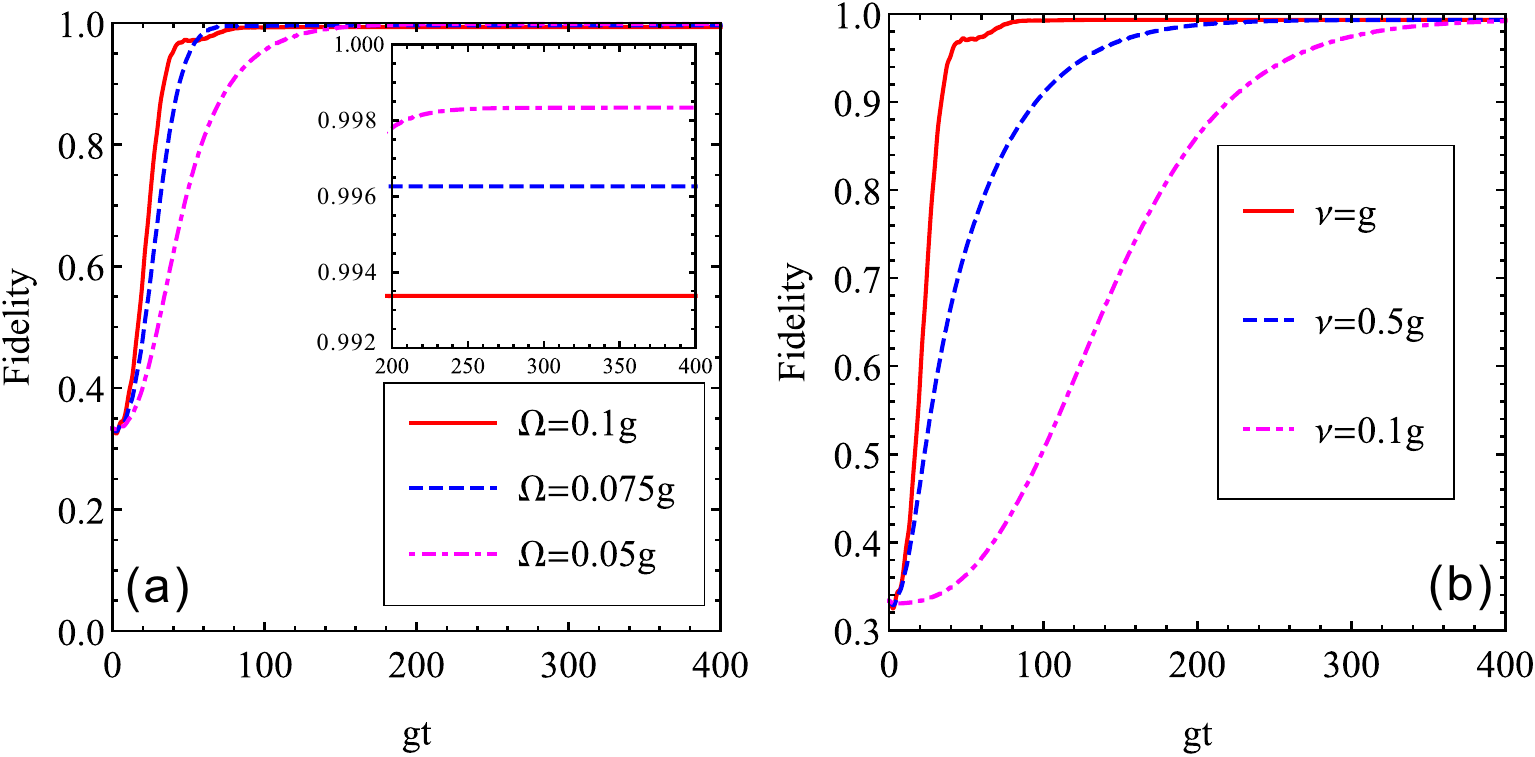}
	\caption{The time evolution of the fidelity for different
		(a) coupling parameter $\Omega$ with $\Delta=0$, $\nu=g$, and $K=0.2$
		and (b) couping strengths with
		$\nu=g$, $K=0.2$ (red curve), $\nu=0.5g$, $K=1$ (blue curve), and $\nu=0.1g$, $K=2$ (magenta curve)
		with $\Delta=0$ and $\Omega=0.1g$.
		\label{f3-2}}
\end{figure}

In Fig. \ref{f3-1}, we plot the time evolution of the fidelity and the control fields for the initial state $|f,g,g,0,0,0\rangle_{1,2,3,a_1,a_2,b}$.
It shows the control fields steer the system gradually to the W state.
The system is in the steady W state and the fidelity reaches $0.993$ finally.
In fact, the fidelity is limited by the parameter $\Omega /g$.
Because the dark state of the system is approximated to a W state, i.e., the fidelity increases
as the parameter $\Omega/g$ decreases.
Figure \ref{f3-2}(a) illustrates the time evolution of the fidelity for different parameters $\Omega /g$.
Expect for $\Omega$, we also explore the influence of the coupling strength $\nu$ on the fidelity.
In Fig. \ref{f3-2}(b), we plot the time evolution of the fidelity for different
coupling strengths $\nu$.
The fidelity can reach the maximum ($0.993$) under different coupling strengths.
The evolution time increases as the coupling strength $\nu$ decreases.
It implies that a high fidelity W state can be generated, even when the coupling strength between
cavity and fiber is very weak.

%%Fig.2

\section{Lyapunov control in open systems}\label{s3}

\subsection{The influence of atomic spontaneous emission}

In this section, we consider the influence of atomic spontaneous emission on the fidelity.
The dynamical evolution of the system is described by the master equation, which is expressed by
\begin{equation}
\dot{\rho}=-\mathrm{i}[H_{\mathrm{I}}+\sum_k f_k(t)H_k,\rho]+\mathcal{L}(\rho),
\end{equation}
where $\mathcal{L}(\rho)$ is the Lindblad operator defined by
$\mathcal{L}(\rho)=\sum_{j=1,2,3}\sum_{i=g,f}\gamma_{ij}(\sigma_{ij}\rho\sigma_{ij}^{\dag}
-\frac{1}{2}\rho\sigma_{ij}^{\dag}\sigma_{ij}-\frac{1}{2}\sigma_{ij}^{\dag}\sigma_{ij}\rho)$.
$\sigma_{ij}=|i_j\rangle\langle e_j|$ and $\gamma_{ij}$ is the emission rate from $|e_j\rangle$ to $|i_j\rangle$.
For simplicity, we assume $\gamma_{ij}=\gamma$.
The Lyapunov function is defined as $V=1-\mathrm{Tr}(\rho_{\mathrm{T}} \rho)$ and the time derivative of the Lyapunov function
can be calculated as
\begin{equation}
\dot{V}=-\mathrm{Tr}(\rho_{\mathrm{T}}\mathcal{L}(\rho))-\sum_kf_k(t)\mathrm{Tr}(-\mathrm{i}\rho_{\mathrm{T}}[H_k,\rho]).
\end{equation}
For guaranteeing $\dot{V}\le 0$, the control fields can be designed as
\begin{equation}
f_1(t)=-\frac{\mathrm{Tr}(\rho_{\mathrm{T}}\mathcal{L}(\rho))}{\mathrm{Tr}(-\mathrm{i}\rho_{\mathrm{T}}[H_1,\rho])},\quad
f_k(t)=\mathrm{Tr}(-\mathrm{i}\rho_{\mathrm{T}}[H_k,\rho]),\quad k=2,3.
\end{equation}
Here, we choose the control field $f_1(t)$ to eliminate the influence of atomic spontaneous emission.
The control fields $f_2(t)$ and $f_3(t)$ ensure the time derivative of Lyapunov function $\dot{V}\le 0$.

%Fig.3
\begin{figure}
	\centering
	\includegraphics[width=0.8\textwidth]{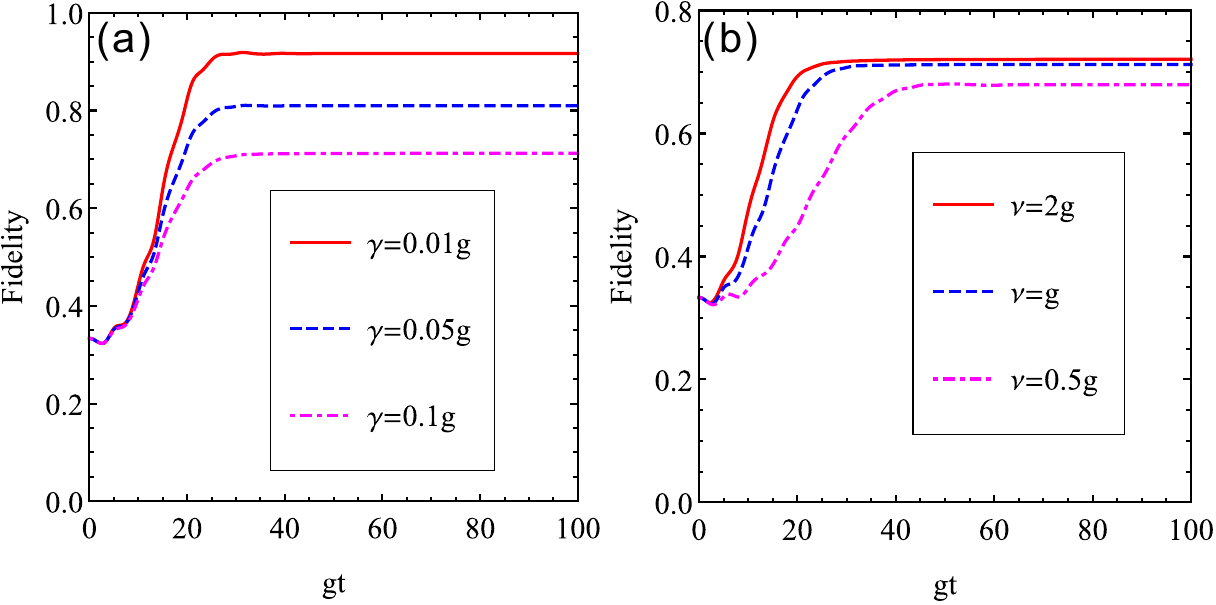}
	\caption{The time evolution of the fidelity for different
		(a) atomic spontaneous emission rates with $\nu=g$ 
		and (b) coupling strengths with $\gamma=0.1g$. Other common parameters: $\Delta=0$ and $\Omega=0.1g$.
		\label{f4a1}}
\end{figure}

Figure \ref{f4a1}(a) illustrates the time evolution of the fidelity for different
atomic spontaneous emission rates.
The fidelity decreases as atomic spontaneous rates increase.
When atomic spontaneous emission rate is small, the fidelity can reach a high value ($0.91$).
In the case of large atomic spontaneous emission rate, however, the fidelity can only reach $0.71$.
The coupling strength $\nu$ is an important parameter of the system, which influences the interaction between two
cavities.
We plot the time evolution of the fidelity for different coupling strengths $\nu$ in
Fig. \ref{f4a1}(b).
It shows that the fidelity increases as the coupling strength increases.
It is because the large coupling strength speeds up the evolution of the system, which suppresses atomic
spontaneous emission.
However, the fidelity almost remains unchanged with further increase of coupling strength $\nu$, when $\nu \ge 2 g$.
That because atomic spontaneous emission plays a dominant role in the dynamical evolution of the system.

As mentioned above, the fidelity is low in the case of large dissipation.
To further improve the fidelity, the single qubit operation $\sigma_j$ will be
performed on atom $j$, where $\sigma_j=|f_j\rangle \langle e^{'}_j|+|e^{'}_j\rangle \langle f_j|$ ($j=1,2,3$).
Here $|e^{'}_j\rangle$ and $|g^{'}_j\rangle$ are the auxiliary excited and ground state of atom $j$ with
energies $\omega_{e_j}^{'}$ and $\omega_{g_j}^{'}$, respectively.
The transition between $|e^{'}_1\rangle$ and $|g^{'}_1\rangle$ is coupled to the cavity mode $b_1^{'}$ of
cavity $1$ with coupling strength $g_1^{'}$.
The transition between $|e^{'}_{2(3)}\rangle$ and $|g^{'}_{2(3)}\rangle$ is coupled to the cavity mode
$b_2^{'}$ of cavity $2$ with coupling strength $g_2^{'}$ ($g_3^{'}$).
When the system reaches the steady state at $t_0$, the state $|f_j\rangle$ is driven to
the auxiliary excited state $|e^{'}_j\rangle$ by the classical field $j^{'}$ with strength $\Omega_j^{'}$ and
frequency $\omega_{l_j}^{'}$.
The auxiliary excited state $|e^{'}_j\rangle$ will decay to
the auxiliary ground state $|g^{'}_j\rangle$ with emitting a photon into cavity $j$, which need to be
detected \cite{Li:17}.
The driving Hamiltonian is given by
\begin{eqnarray}
H^{'}&=&\sum_{j=1}^3(\omega_{e_j}^{'}|e_j^{'}\rangle\langle e_j^{'}|+\omega_{g_j}^{'}|g_j^{'}\rangle\langle g_j^{'}|
+\omega_{f_j}|f_j\rangle\langle f_j|) \nonumber \\
&&+\sum_{i=1,2}\omega_{b_i}b_i^{\dag}b_i
+\sum_{j=1}^3(\Omega_j^{'}|e_j^{'}\rangle\langle f_j |\mathrm{e}^{-\mathrm{i} \omega_{l_j}^{'} t}+\mathrm{H.c.}) \nonumber \\
&&+(g_1^{'}|e_1^{'}\rangle\langle g_1^{'}|b_1+g_2^{'}|e_2^{'}\rangle\langle g_2^{'}|b_2
+g_3^{'}|e_3^{'}\rangle\langle g_3^{'}|b_2+\mathrm{H.c.}),
\end{eqnarray}
where $\omega_{b_i}$ is the frequency of the cavity mode $b_i$.
After the driving Hamiltonian, the state $|f_j\rangle$ will evolve to state $|g^{'}_j\rangle$.
In the case of large detunings, the effective driving Hamiltonian in the interaction picture can be obtained as \cite{doi:10.1139/p07-060}
\begin{eqnarray}
H^{'}_{\mathrm{eff}}&=&\sum_{j=1}^3\frac{\Omega_j^{'2}}{\Delta_{l_j}}|f_j\rangle\langle f_j|
+\frac{g_j^{'2}}{\Delta_{c_j}}|g_j^{'}\rangle\langle g_j^{'}|b_j^{\dag}b_j \nonumber\\
&&+\frac{g_j^{'}\Omega_j^{'}}{2}(\frac{1}{\Delta_{l_j}}+\frac{1}{\Delta_{c_j}})
[|f_j\rangle\langle g_j^{'}|b_j\mathrm{e}^{\mathrm{i}(\Delta_{l_j}-\Delta_{c_j})t}+\mathrm{H.c.}],
\end{eqnarray}
where $\Delta_{l_j}^{'}=\omega_{l_j}-(\omega_{e^{'}_j}-\omega_{f_j})$,
$\Delta_{c_1}=\omega_{b_1}-(\omega_{e_1}^{'}-\omega_{g_1}^{'})$,
$\Delta_{c_2}=\omega_{b_2}-(\omega_{e_2}^{'}-\omega_{g_2}^{'})$, and
$\Delta_{c_3}=\omega_{b_2}-(\omega_{e_3}^{'}-\omega_{g_3}^{'})$.
This process is described by the following master equation
\begin{equation}
\dot{\rho}=-\mathrm{i}[H^{'}_{\mathrm{eff}},\rho]+\kappa^{'}(D\rho D^{\dag}-\frac{1}{2}\rho D^{\dag}D-\frac{1}{2}D^{\dag}D\rho),
\end{equation}
where $D=\frac{1}{\sqrt{2}}(b_1+b_2)d^{\dag}$, $d$ is the annihilation operator of a detector mode,
and $\kappa^{'}$ is the decay rate of the cavities.
The master equation describes an irreversible detection process.
By detecting the photon at $t$ ($t>t_0$), the system is projected to the subspace of the auxiliary states.
Correspondingly, the density matrix is expressed by $\rho^{'}=P\rho P^{-1}\left[\mathrm{Tr}(P\rho P^{-1})\right]^{-\frac{1}{2}}$ with $P=|1 \rangle_d{}_d\langle 1|$.

\begin{figure}
	\centering
	\includegraphics[width=0.4\textwidth]{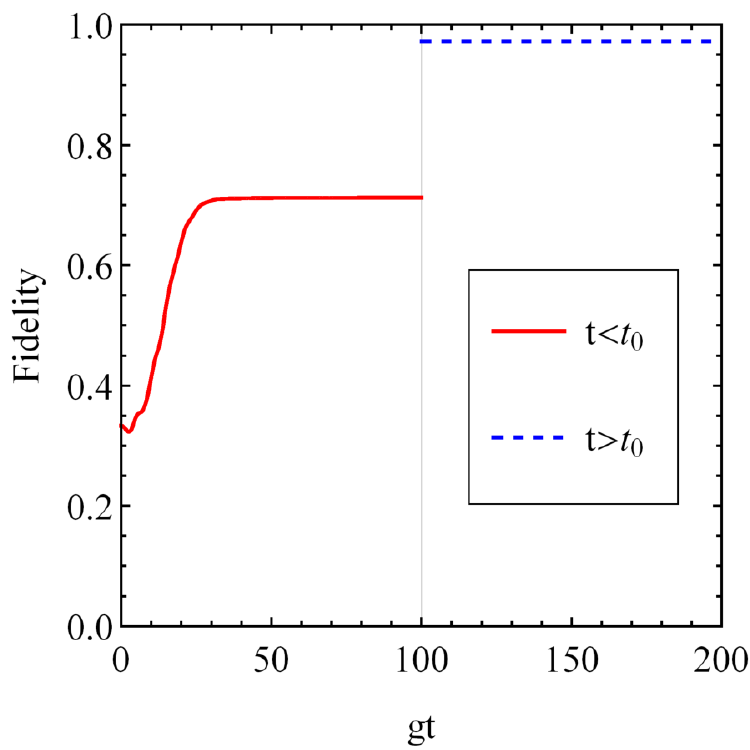}
	\caption{The time evolution of the fidelity before and after detection. Other parameters:
		$\Delta=0$, $\Omega=0.1g$, $\gamma=0.1g$, $\kappa ^{'}=0.1g$, $\Delta_{c_j}=\Delta_{l_j}=10g$,
		and $\Omega_j^{'}=g_j^{'}=g$ ($j=1,2,3)$.\label{f4a3}}
\end{figure}

In Fig. \ref{f4a3}, we plot the time evolution of the fidelity
before and after detection (detection time $t_0=100gt$).
It shows that the fidelity is improved from $0.71$ to $0.97$ in the case of large atomic spontaneous emission rate.

%fig.4

\subsection{The influence of cavity and fiber decay}
In this section, we consider the influence of cavity and fiber decay.
The dynamical evolution of the system is described by the master equation, which is expressed by
\begin{equation}
\dot{\rho}=-\mathrm{i}[H_{\mathrm{I}}+\sum_k f_k(t)H_k,\rho]+\mathcal{L}_{\mathrm{c}}(\rho),
\end{equation}
where $\mathcal{L}_{\mathrm{c}}(\rho)=\sum_{i=a_1,a_2,b}\gamma_{i}(i\rho i^\dag-\frac{1}{2}\rho i^\dag i-\frac{1}{2}i^\dag i\rho)$
and $\gamma_i$ is the decay rate of the cavity (fiber) mode.
For convenience, we set $\gamma_i=\gamma$.
With the similar derivation procedures, the control fields can be designed as

\begin{equation}
f_1(t)=-\frac{\mathrm{Tr}(\rho_{\mathrm{T}}\mathcal{L}^{'}(\rho))}{\mathrm{Tr}(-\mathrm{i}\rho_{\mathrm{T}}[H_1,\rho])}, \quad
f_k(t)=\mathrm{Tr}(-\mathrm{i}\rho_{\mathrm{T}}[H_k,\rho]),\quad k=2,3.
\end{equation}

%%Fig. 5
\begin{figure}[hthp]
	\centering
	\includegraphics[width=0.8\textwidth]{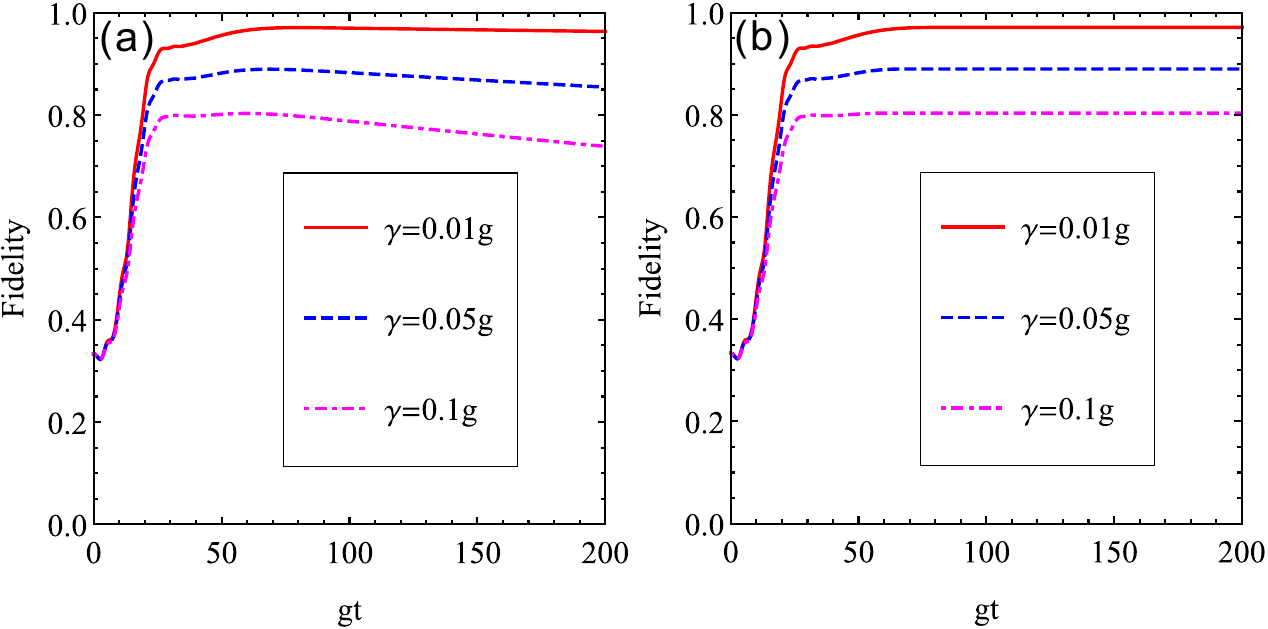}
	\caption{The time evolution of the fidelity for different
		cavity and fiber decay rates.
		Other parameters:
		$\Delta=0$, $\Omega=0.1g$, and $\nu=g$.\label{f4b1}}
\end{figure}

%%Fig. 6
\begin{figure}
	\centering
	\includegraphics[width=0.4\textwidth]{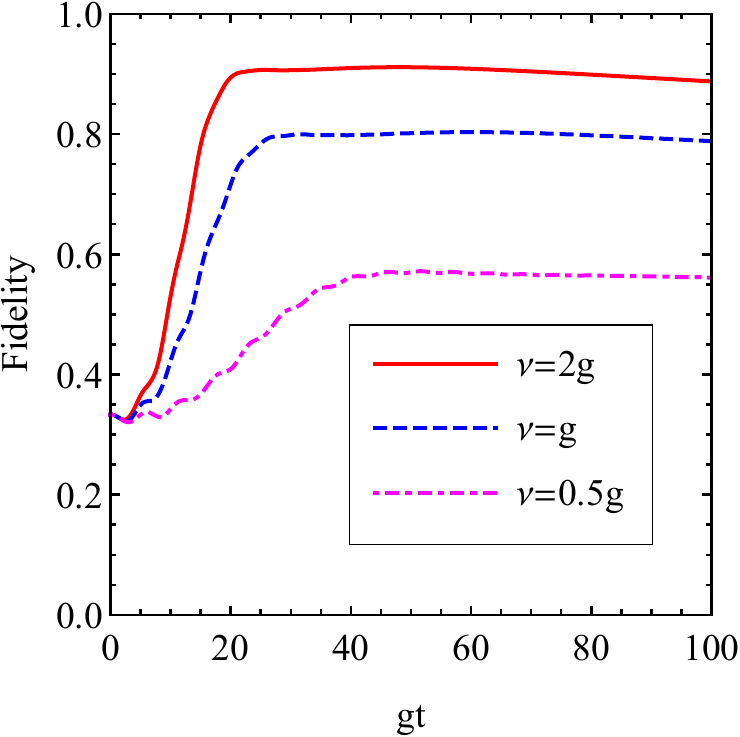}
	\caption{The time evolution of the fidelity for different coupling strengths.
		Other parameters:
		$\Delta=0$, $\Omega=0.1g$, and $\gamma=0.1g$.\label{f4b2}}
\end{figure}

Figure \ref{f4b1} illustrates the time evolution of the fidelity for different
cavity and fiber rates.
It shows that the fidelity decreases as cavity and fiber decay rates increase.
The fidelity first increases to a maximum and then gradually decreases.
However, the fidelity can maintain the maximum by switch off all the control fields and classical fields
(Fig. \ref{f4b1}(b)).
In Fig. \ref{f4b2}, we plot the time evolution of the fidelity for different
coupling strengths $\nu$.
It shows that the maximum of fidelity increases as the coupling strength increases.
When the coupling strength is large, the fidelity reaches the maximum in a short time.

In the case of large cavity and fiber decay rate, however, the fidelity can only reach $0.79$.
Hence, we introduce quantum feedback based on quantum-jump-detection to improve the fidelity \cite{PhysRevA.49.2133}.
The dynamical evolution of the system satisfies the master equation with quantum feedback, which is given by
\begin{equation}
\dot{\rho}=-\mathrm{i}[H_{\mathrm{I}}+\sum_k f_k(t)H_k,\rho]+\eta\mathcal{L}_{\mathrm{f}}(\rho)+(1-\eta)\mathcal{L}_{\mathrm{c}}(\rho),
\end{equation}
where $\mathcal{L}_{\mathrm{f}}(\rho)=
\sum_{i=a_1,a_2,b}\gamma_{i}(Fi\rho i^\dag F^\dag-\frac{1}{2}\rho i^\dag i-\frac{1}{2}i^\dag i\rho)$
and $\eta$ is the efficiency of detectors.
$F=\exp(\mathrm{i} H_{\mathrm{f}})$ is feedback operator, where $H_{\mathrm{f}}=\frac{\pi}{2}(|f_1\rangle\langle g_1|+|g_1\rangle\langle f_1|)$.
we use the control fields in closed systems.

%%Fig. 7
\begin{figure}
	\centering
	\includegraphics[width=0.8\textwidth]{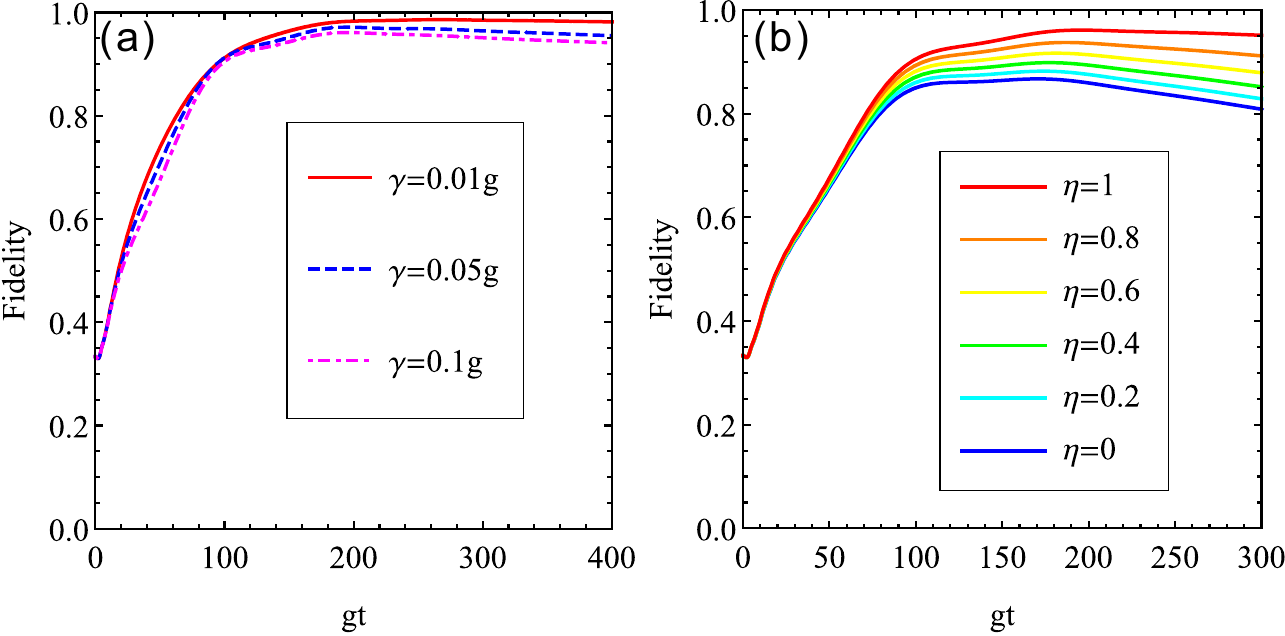}
	\caption{The time evolution of the fidelity with quantum feedback for different (a) cavity and fiber decay rates
		with $\eta=1$ and
		(b) efficiencies of detectors with  $\gamma=0.1g$.
		Other common parameters:
		$\Delta=0$, $\Omega=0.1g$, and $\nu=g$.\label{f4b3}}
\end{figure}

In Fig. \ref{f4b3}(a), we plot the time evolution of the fidelity with quantum feedback for different cavity and fiber decay rates.
The results show that the maximum of the fidelity reaches $0.96$ in the case of large decay rates ($\gamma=0.1g$).
Compared with the case without quantum feedback, the fidelity increases by $0.16$.
As quantum feedback is dependent on the efficiency of detectors,
we plot the time evolution of the fidelity as a function of the scaled time $gt$ for different
efficiencies of detectors in Fig. \ref{f4b3}(b).
It shows that the maximum of the fidelity decreases as the efficiency of detectors decreases.
when the efficiency of detectors is greater than 0.4, the fidelity can reach 0.9.

\section{Time-optimal Lyapunov control}\label{s4}
In this section, we consider time-optimal Lyapunov control.
We optimize the control fields to speed up the evolution of the system, which can be achieve
by increasing the time derivative of the Lyapunov function $|\dot{V}|\propto |f_k(t)|$.
However, too strong control fields may disturb the quantum system and lead to invalidation of the physical model.
In real systems, the control fields are under the constraints of energy and strength.

In the case of the power constraint $W(t)=\sum_{k}f_k(t)^2\le W_{\mathrm{max}}$,
the Lagrange multiplier method can be used to determine the control fields which minimize $\dot{V}$ ($\dot{V}<0$) \cite{PhysRevA.86.022321}.
According to the Lagrange function $\mathcal{L}=-\sum_k f_k(t) T_k+\lambda \left[ \sum_k f_k^2(t)-W(t) \right ]$
($\lambda$ is the Lagrange multiplier),
we calculate the gradient of Lagrange function and obtain
\begin{subequations}
\begin{eqnarray}
\frac{\partial}{\partial f_k(t)}L&=&-T_k+2\lambda f_k(t)=0,\\
\frac{\partial}{\partial \lambda}L&=&\sum_k f_k(t)^2-W(t)=0,\\
f_k(t)&=&\frac{\sqrt{W(t)}T_k}{\sqrt{\sum_k T_k^2}}.
\end{eqnarray}
\end{subequations}
Thus, the control fields can be designed as follows:
\begin{equation}
f_k(t)=\left\{
\begin{array}{ll}
\frac{\sqrt{W_{\mathrm{max}}}T_k}{\sqrt{\sum_k T_k^2}} &   \sum_k T_k^2 \ne 0, \\
0 & \sum_k T_k^2 = 0.
\end{array}
\right.
\end{equation}

In the case of the strength constraint $|f_k(t)|\le S$ ($S$ is the maximum strength of each control field), 
we can design the optimized control fields according to Eq. (\ref{18}) as follows
\begin{equation}
f_k(t)=\left\{
\begin{array}{ll}
S &   T_k>0, \\
-S &   T_k<0, \\
0 &  T_k=0. \\
\end{array}
\right.
\end{equation}

%%Fig. 8
\begin{figure}
	\centering
	\includegraphics[width=0.4\textwidth]{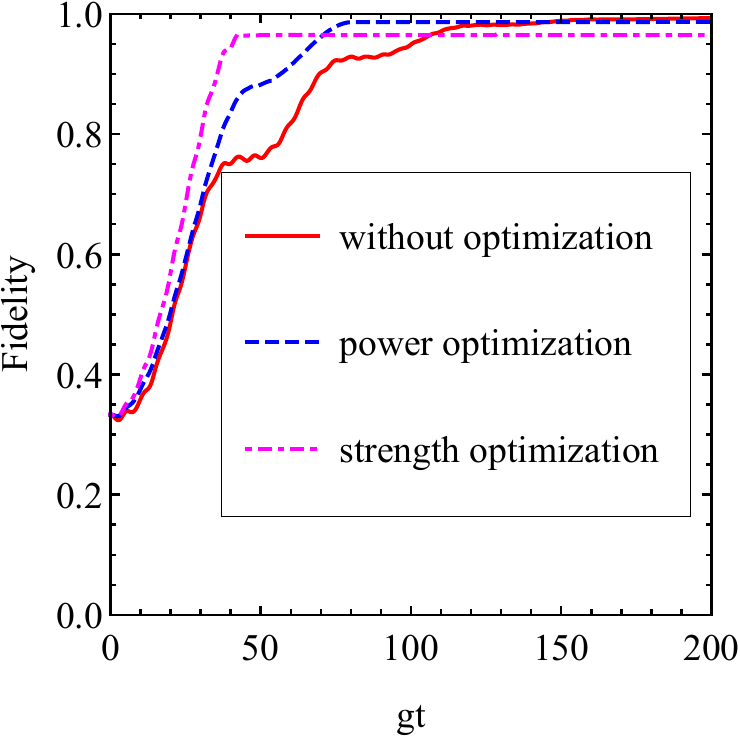}
	\caption{The time evolution of the fidelity with time-optimal Lyapunov control in closed systems.
		Other parameters: $\Delta=0$, $\Omega=0.1g$, $\nu=g$, $W_{\mathrm{max}}=0.002$, and $S=0.038$.\label{f51}}
\end{figure}

%Fig. 9
\begin{figure}[hthp]
	\centering
	\includegraphics[width=0.8\textwidth]{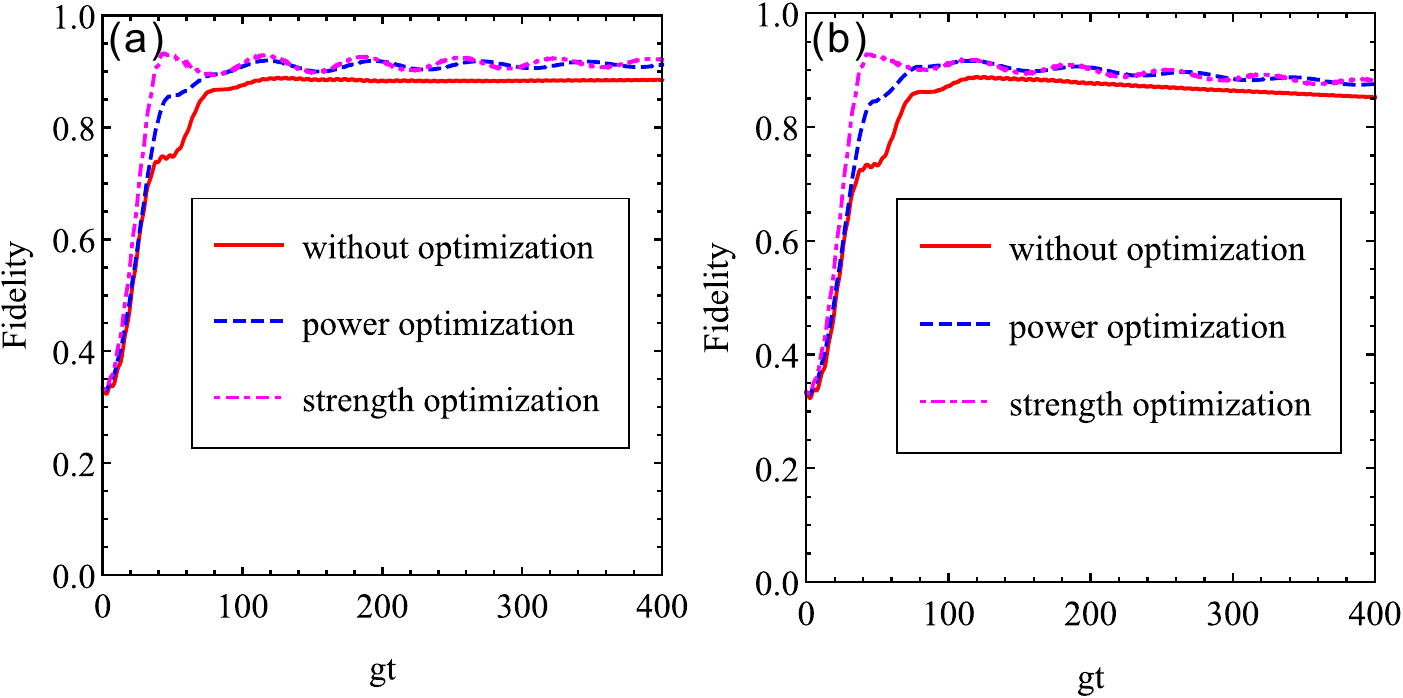}
	\caption{The time evolution of the fidelity with time-optimal Lyapunov control in open systems. Other parameters:
		$\Delta=0$, $\Omega=0.1g$, $\nu=g$, $\gamma=0.01g$, $W_{\mathrm{max}}=0.002$, and $S=0.038$.\label{f52}}
\end{figure}

In Fig. \ref{f51}, we plot the time evolution of the fidelity with time-optimal Lyapunov control in closed systems.
The results show that two time-optimal designs can shorten the time required to reach the target state.
Especially, the evolution time is the shortest with the design under the strength constraint.
In closed systems, however, the two time-optimal designs lead to a loss of fidelity due to the additional
constraints on the control fields.
In Fig. \ref{f52}, we plot the time evolution of the fidelity with time-optimal Lyapunov control in open systems:
(a) atomic spontaneous emission and (b) cavity and fiber decay.
In open systems, the fidelity with two optimal designs is obviously higher than that without optimization,
which is different from the case in closed systems.
It implies that time-optimal Lyapunov control is an effective method against decoherence.

\section{Conclusion}\label{s5}
In summary, we have investigated the system formed by two distant cavities connected by an optical fiber,
one of which contains an atom and the other contains two atoms.
We generate W state of atoms via Lyapunov control in closed and open systems.
In closed systems, the fidelity increases as the coupling strength $\Omega$ decreases and
a high fidelity W state can be generated in the case of weak coupling between the cavities and fiber.
In open systems, the fidelity decreases with the decay rates increase and the fidelity is low in
the case of large decay rates.
Hence, we propose two schemes to improve the fidelity in open systems.
In the presence of atomic spontaneous emission, we drive each atom to an auxiliary exited state, which
decays to the auxiliary ground state with emitting a photon into the cavity.
The fidelity can be improved by detecting the photon leaking from the cavities.
In the case of cavity and fiber decay, we use quantum feedback based on quantum-jump-detection to improve
the fidelity.
The results show the fidelity is improved greatly by using the two schemes.
In addition, we consider time-optimal Lyapunov control with two constraints (power and strength).
It shows that the time-optimal Lyapunov control can speed up the evolution of the system and improve the
fidelity in open systems.

\section*{Funding}
National Natural Science Foundation of China (NSFC) (11474077, and 11675046), Program for Innovation Research of Science in Harbin Institute of Technology (A201411, and A201412), the Fundamental Research Funds for the Central
Universities (AUGA5710056414), Natural Science Foundation of Heilongjiang Province of
China. (A201303), and Postdoctoral Scientific Research Developmental Fund of Heilongjiang
Province (LBH-Q15060).

\end{document}